\documentclass[aps,prl,twocolumn,superscriptaddress]{revtex4}
\usepackage{graphicx}

\begin{document}


\title{Electronically smectic-like phase
in a nearly half-doped manganite}

\author{F. Ye}
\affiliation{Center for Neutron Scattering,
Oak Ridge National Laboratory, Oak Ridge, Tennessee 37831-6393,
USA}

\author{J. A. Fernandez-Baca}
\affiliation{Center for Neutron Scattering,
Oak Ridge National Laboratory, Oak Ridge, Tennessee 37831-6393,
USA}
\affiliation{Department of Physics and Astronomy,
The University of Tennessee, Knoxville, Tennessee 37996-1200, USA}

\author{Pengcheng Dai}
\affiliation{Department of Physics and Astronomy,
The University of Tennessee, Knoxville, Tennessee 37996-1200, USA}
\affiliation{Center for Neutron Scattering,
Oak Ridge National Laboratory, Oak Ridge, Tennessee 37831-6393,
USA}

\author{J. W. Lynn}
\affiliation{NIST Center for Neutron Research,
Gaithersburg, Maryland, 20899, USA}

\author{H.~Kawano-Furukawa}
\affiliation{Department of Physics,
Ochanomizu University, Bunkyo-ku, Tokyo 112-8610, Japan
}

\author{H. Yoshizawa}
\affiliation{Neutron Science Laboratory, ISSP,
University of Tokyo, Tokai, Ibaraki, 319-1106, Japan
}

\author{Y. Tomioka}
\affiliation{Correlated Electron Research Center (CERC), 
Tsukuba 305-0046, Japan}

\author{Y. Tokura}
\affiliation{Correlated Electron Research Center (CERC),
Tsukuba 305-0046, Japan}
\affiliation{Department of Applied Physics, University of Tokyo,
Tokyo 113-8656, Japan}

\date{\today}

\begin{abstract}
We use neutron scattering to study the spin and charge/orbital
ordering (CO-OO) in the nearly half-doped perovskite manganite $\rm
Pr_{0.55}(Ca_{0.8}Sr_{0.2})_{0.45}MnO_3$ (PCSMO). On cooling from
room temperature, PCSMO first enters into a CO-OO state below
$T_{CO}$ and then becomes
a CE-type long-range ordered antiferromagnet below $T_N$. 
At temperatures above $T_N$ but 
below $T_{CO}$ ($T_N<T<T_{CO}$), the spins in
PCSMO form highly anisotropic smectic liquid-crystal-like
texture with ferromagnetic (FM) quasi-long-range ordered
one-dimensional zigzag chains weakly coupled
antiferromagnetically. Such a magnetic smectic-like phase results directly from the spin-orbit
interaction and demonstrates the presence of textured `electronic soft'
phases in doped Mott insulators.
\end{abstract}



\maketitle
 
The manganese oxides with general composition ${\rm R}_{1-x}{\rm
A}_x{\rm MnO}_3$ (where R and A are rare- and alkaline-earth
ions) have been actively investigated over the past decades
because of the colossal magnetoresistance (CMR) effect observed
around $x=0.30$. The fascinating properties of these materials
have been attributed to the electronic complexity arising from
the strong competition between charge, orbital, lattice and
magnetic degrees of freedom \cite{dagotto05,salamon01}.  
Of particular interest is the case of the half-doped ($x=0.5$) 
manganites.
When cooling from room temperature, they form a structure that has been
described as a ``charge and orbital ordered''
(CO-OO) state, where Mn$^{3+}$ and $\rm Mn^{4+}$ ions are
arranged alternatively in the form of checkerboard-like order
below $T_{CO}$ \cite{wollan55,goodenough55}.  On further cooling,
the spins order into the CE-type magnetic structure below $T_N$
(Fig.\ 1(a)). In the intermediate temperature region
above $T_N$ but below $T_{CO}$ ($T_N<T<T_{CO}$), there is a
coexistence of incommensurate and inhomogeneous magnetic and
CO-OO ordered states \cite{kajimoto98}. The nature of such
coexistence is still unclear and it has been recently suggested 
that it is related to the emergence
of new electronic soft phases \cite{milward05,dagotto05}.
In this Letter, we present comprehensive neutron scattering
studies on a nearly half-doped perovskite manganite $\rm
Pr_{0.55}(Ca_{0.8}Sr_{0.2})_{0.45}MnO_3$ (PCSMO) focusing in
the purported ``electronically soft'' phase regime of $T_N<T<T_{CO}$. We
choose PCSMO because this material has well separated CO-OO and
AF transitions ideally suited for our investigation
\cite{tomioka02}.  While the CO-OO fluctuations in PCSMO are
spatially isotropic as expected, the AF spin fluctuations are 
highly anisotropic and indicative of the formation of 
quasi-long-range FM ordered 1D spin chains 
that are weakly coupled antiferromagnetically.
The anisotropic and low-dimensional character of the
spin fluctuations is counterintuitive in a
conventional paramagnetic state, as PCSMO is essentially cubic.
We argue that these features are 
due to the anisotropy of the short-range Double 
Exchange interactions as proposed by Van den Brink et al and 
Solovyev and Terakura \cite{brink99,solovyev99}, and that the complex coexistence 
of the magnetic and charge/orbital fluctuations in this regime arises 
from the magnetic exchange and the orbital degeneracy of the Mn ions.
Our results highlight the role of the orbital physics in determinig the 
electronic properties of the half-doped manganites and support the idea that
the charge/orbital and magnetic fluctuations have the same magnetic origin.
The observed highly anisotropic spin texture is reminiscent of the smectic phases
\cite{smectic} predicted in the doped Mott insulators
\cite{kivelson98} and is the first example
of the textured ``electronically soft'' phases in the CMR
manganites \cite{milward05}. 


\begin{figure}[ht!]
\includegraphics[width=3.0in]{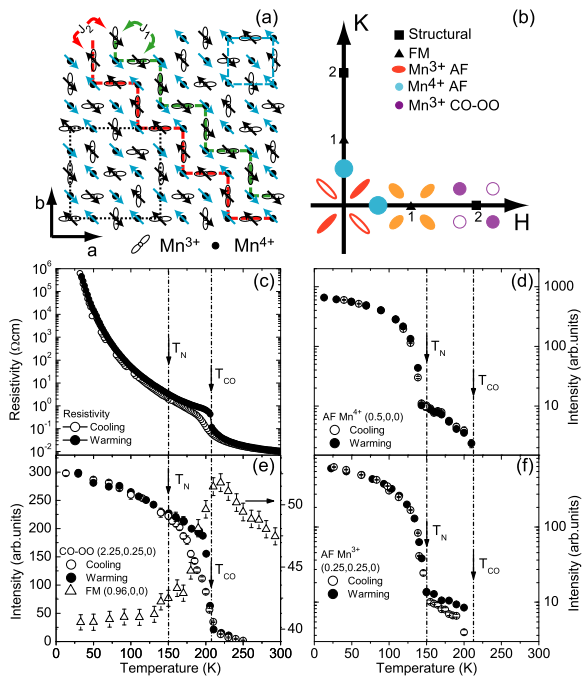}
\caption{\label{fig:fig1}
Panel (a) shows schematic diagram of the CE-type structure in the
nearly half doped perovskite manganites in the cubic setting. The
zigzag chains formed by alternating $\rm Mn^{3+}$ and $\rm
Mn^{4+}$ spins are coupled antiferromagnetically, the coupling
within the chain is ferromagnetic. The short-dashed line denotes
the periodicity of the Mn$^{3+}$ orbital and magnetic unit cell,
long-dashed line shows the Mn$^{4+}$ magnetic unit cell. Panel
(b) depicts the corresponding superlattice peaks in the
reciprocal space (solid symbols). Open symbols indicate superlattice peaks
that are forbidden in a single-domain crystal but are observed
because of twining.  Bottom: Temperature dependence of
(c) resistivity; (d) AF peak intensity from (0.5,0,0); (e) CO-OO
peak intensity from (2.25,0.25,0) and FM short-range fluctuations
from (0.96,0,0);  and (f) AF peak intensity from (0.25,0.25,0).
In panels (c-f) open and solid symbols represent cooling and
warming, respectively.} 
\end{figure}

We grew a single crystal of PCSMO using the floating 
zone method \cite{tomioka02}.  
PCSMO has the $Pbnm$ symmetry in the low-temperature orthorhombic 
phase ($a \approx b\approx
3.85$~\AA\ and $c \approx 3.79$~\AA\ below $T_{CO}$), slightly distorted from the
cubic lattice. 
The sample is naturally twinned due to the small orthorhombicity, 
the twinning domains are 90 degree rotated with respect to 
each other in the scattering plane.
For simplicity, we use a pseudo-cubic unit cell
with lattice parameters of $a \approx b\approx c \approx 3.84
$~\AA, and all the wavevectors below refer to the cubic notation. 
The momentum transfers $\vec{q}=(q_x, q_y, q_z)$ in units
of $\AA^{-1}$ are at positions $(h,k,l)=(q_x a/2\pi,q_y
a/2\pi,q_z a/2\pi)$ in reciprocal lattice units (rlu). The sample
was aligned to allow the wavevector in the form of $(h,k,0)$
accessible in the horizontal scattering plane.
Our neutron scattering experiments were carried out using thermal
neutron triple-axis spectrometers at the High-Flux Isotope
Reactor (HFIR), Oak Ridge National Laboratory, the NIST Center
for Neutron Research, and the JRR--3 reactor JAERI, Tokai, Japan.
The neutron energy was fixed at $E=14.7~meV$ using pyrolytic
graphite crystals as monochromators, analyzers and filters.

Upon cooling PCSMO undergoes a series of transitions. It first
enters the CO-OO state below $T_{CO} \sim$ 210~K and becomes a
long-range antiferromagnet below $T_N \sim$ 150~K with a
CE-type spin arrangement \cite{wollan55}.
Neutron diffraction is not sensitive to the ordering of orbitals
but it can measure the superlattices associated with the lattice
distortions caused by such order. In a twinned crystal, the
characteristic propagation wavevectors of the CO-OO structure are
$\vec{q}=(1/4,1/4,0)$ (which is also the propagation wave vector
for the network of $\rm Mn^{3+}$ spins), and $\vec{q}=(1/2,0,0)$ is the
propagation wave vector for the $\rm Mn^{4+}$
spins network \cite{jirak85}.  Experimentally the intensity of the
scattering at a wave vector $\vec{Q}=(\vec{q}+\vec{\tau})$, where
$\vec{\tau}$ is a vector of the reciprocal lattice and $\vec{q}$
is the propagation wave vector, provides a measurement of the
order parameter for such structure.  Since the neutron scattering
cross section is proportional to the square of the magnetic form
factor, which decreases rapidly at large scattering wavevectors,
one can choose ``large'' wave vectors $\vec{Q}$ to probe the CO-OO
and ``small'' $\vec{Q}$'s to probe the magnetic ordering. In our
neutron diffraction experiments, we used $\vec{Q}=(2.25,0.25,0)$,
$(0.25,0.25,0)$ and $(0.96,0,0)$ to measure the CO-OO, AF and FM
fluctuations, respectively. We also used $\vec{Q}=(0.5,0,0)$ to
probe the AF fluctuations related to the $\rm Mn^{4+}$ spins
(Fig.\ 1(b)).

Figures 1(c)-(f) show the temperature dependence of
the resistivity \cite{tomioka02} and the peak intensities
associated with AF, FM, CO-OO fluctuations as the sample is
cooled in zero magnetic field. The resistivity exhibits
insulating behavior at all temperatures with a distinct feature
around $T_{CO}=210$~K in the $\rho$-$T$ curve. The hysteresis in
the resistivity near $T_{CO}$ is accompanied by a first-order
structural transition from pseudo-cubic to orthorhombic phase
causing a small splitting of the (2,0,0) lattice peak. The CO-OO peak
intensity in Fig.\ 1(e) also shows an abrupt change
near $T_{CO}$ with strong hysteresis. The clear correlation
between the resistivity and the CO-OO peak intensity indicates
that the transport properties of these materials are mostly
governed by the localization and delocalization of charge carries
\cite{dai00b,adams00}. PCSMO does not exhibit long-range FM
order in zero magnetic field but develops short-range FM spin
fluctuations above $T_N$. These fluctuations increase with
decreasing temperature, peak at $T_{CO}$, and are greatly
suppressed at low temperatures (Fig.\ 1(e)). The
most interesting findings come from the temperature dependence of
the AF intensities associated with the $\rm Mn^{3+}$and $\rm Mn^{4+}$
spin fluctuations.  As shown in Figs.\ 1(d) and 1(f), weak AF spin
fluctuations appear just below $T_{CO}$ and increase gradually
with cooling. An abrupt change takes place below $T_{N} \sim
150$~K, where the sample enters the long-range AF ordered
pseudo-CE phase. 

\begin{figure}[ht!]
\includegraphics[width=2.9in]{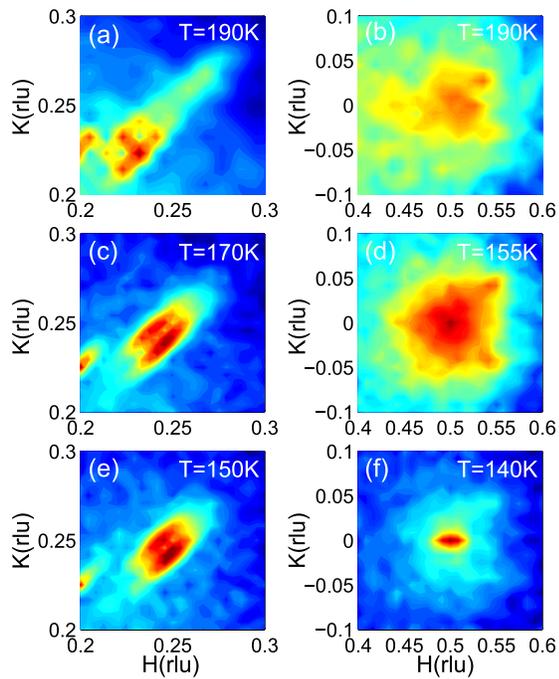}
\caption{\label{fig:fig2} Panels (a), (c), (e) show mesh scans 
around the AF Bragg peaks (0.25,0.25,0) at $T=190$, 170 and 150~K. 
Panels (b), (d), (f) show mesh scans around (0.5,0,0) at
$T=190$, 155 and 140~K.} 
\end{figure}

To probe the characteristics of AF fluctuations in
$T_N<T<T_{CO}$, we performed mesh scans in reciprocal space near
the $\vec{Q}=(0.25,0.25,0)$ and $\vec{Q}=(0.5,0,0)$ AF ordering
positions corresponding to the 
$\rm Mn^{3+}$ and $\rm Mn^{4+}$ spin fluctutations respectively. 
Panels (a), (c) and (e) of 
Fig.\ 2 show the evolution of the AF ordering
associated with the $\rm Mn^{3+}$ fluctuations at $T=190$, 170
and 150~K. These AF fluctuations are clearly anisotropic, with
the profile in the longitudinal $[1,1,0]$ direction being
considerably broader than that in the transverse $[-1,1,0]$
direction. This indicates that the correlation
of spins in the  $[1,1,0]$ direction is much weaker than along 
 $[-1,1,0]$.
The corresponding evolution of the AF
ordering associated with the $\rm Mn^{4+}$ fluctuations, on the
other hand, remains isotropic at all temperatures as illustrated
in panels (b), (d) and (f).
These observations are consistent with the formation of weakly
antiferromagnetically coupled, quasi-long-range FM ordered 1D $\rm Mn^{3+}$ spins
in an isotropic $\rm Mn^{4+}$ spins background.
While our neutron 
scattering measurements allow us to probe the networks of $\rm Mn^{3+}$ and 
$\rm Mn^{4+}$ spins separately, we cannot overlook the fact that these two
types of spins interact strongly via the Double Exchange (DE) process. 
This mechanism allows electron hopping
between neighboring $\rm Mn^{3+}$ and $\rm Mn^{4+}$ if they have parallel spins and
leads to the formation of the zigzag alternating $\rm Mn^{3+}$ 
/ $\rm Mn^{4+}$ chains shown in Fig. 1(a) \cite{brink99}.

Figure 3 shows the profiles of the scattering from AF and
CO-OO fluctuations at representative temperatures on
cooling and warming. Broad  AF fluctuation peaks 
can be measured below $T_{CO}$ (there is no
measurable signal above this temperature). The peak associated
with the $\rm Mn^{4+}$ AF fluctuations is commensurate  with
$\vec{q_c}=(1/2,0,0)$ at all temperatures on cooling and on
warming (Figs.\ 3(a)-(b)) with no hysteresis.  The peaks associated
with the $\rm Mn^{3+}$ AF and CO-OO fluctuations, on the other
hand, are history dependent. They are incommensurate at $T_{CO}$, shift toward
$\vec{q_c}=(1/4,1/4,0)$ upon cooling, and lock into this
commensurate position at $T_N$. Upon warming to $T_{CO}$ the
scattering remains commensurate as summarized in Figs.\ 4(a) and 4(c).
Figure 4(b)  shows the measured longitudinal and
transverse widths of the scattering associated with $\rm Mn^{3+}$ AF fluctuations.
These linewidths decrease upon cooling and the AF 
scattering becomes isotropic and resolution limited when $T\leq T_N$. 
From these linewidths we calculated the correlation lengths \cite{correlation} along the 1D FM
spin chains (intra-chain coupling) and perpendicular to these (inter-chain coupling).
The intra-chain correlation length $\xi_T$  is around
200~\AA\ at 190~K and is only weakly temperature dependent. 
The inter-chain correlation correlation length
$\xi_L$ increases from 33~\AA\ at $T=190$~K to 75~\AA\ on cooling to 150~K. 
This result indicates that the quasi long-range FM 1D spin chains are only weakly 
coupled below $T_{CO}$, and that the interchain coupling becomes stronger 
as $T_N$ is approached. This highly anisotropic spin texture is reminiscent of the 
electronically liquid-crystal-like smectic phases proposed in the doped Mott
insulators \cite{kivelson98}. 
The scattering from the CO-OO fluctuations is isotropic 
\cite{isotropic} with linewidths that decrease upon cooling.
These linewidths, however, remain greater than the instrumental
resolution (Fig.\ 4(d)), even at the lowest temperature.


\begin{figure}[ht!]
\includegraphics[width=3.0in]{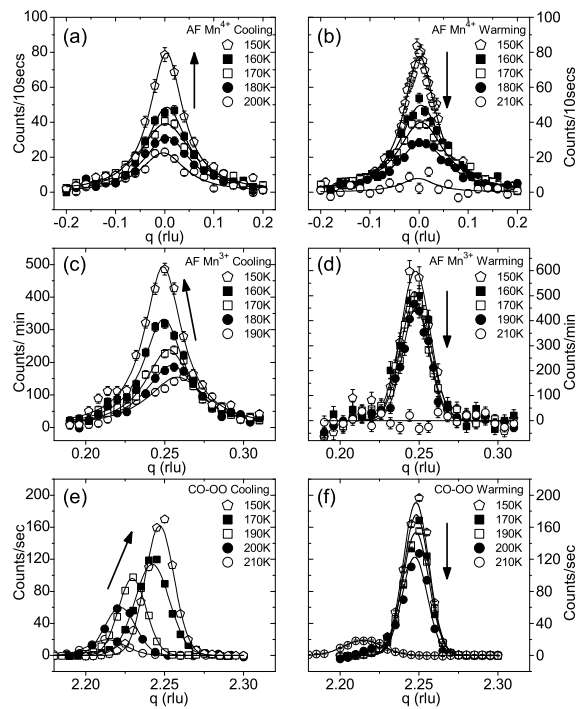}
\caption{\label{fig:fig3} Representative wavevector scans of the
AF component near (0.5,0,0) on (a) cooling and (b) warming;  of
the AF component near (0.25,0.25,0) on (c) cooling and (d)
warming; and CO-OO components near (2.25,0.25,0) on (e) cooling
and (f) warming. The data shown in (c) and (d) have sloping
background subtracted.}
\end{figure}

\begin{figure}[ht!]
\includegraphics[width=3.0in]{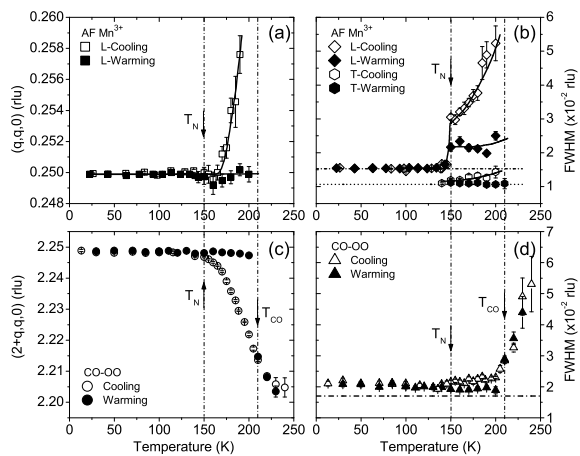}
\caption{\label{fig:fig4}
Temperature dependence of (a) propagation wavevector and (b) FWHM
for $\rm Mn^{3+}$ AF ordering. The corresponding $T$-dependence
of (c) propagation wave vector and (d) FWHM for CO-OO.  
``L'' refers to the [1 1 0] direction and reflects the interchain 
coupling. ``T'' refers to the [-1 1 0] direction and is associated
with the intrachain coupling.  Dashed
lines indicate the instrumental resolution.
}
\end{figure}

The differences among the behaviors described above suggest that
the interplay between the AF and CO-OO fluctuations in the
paramagnetic phase of PCSMO ($T_N\leq T\leq T_{CO}$) is more
complex than expected. The coexistence of AF and CO-OO
fluctuations suggests a strong correlation between them. The
conventional wisdom is that the CO-OO should occur at a higher
temperature, of the order of the covalent bonding energy as
originally proposed by Goodenough \cite{goodenough55},
independent of the magnetic interactions.  However, Goodenough's
conjecture of ordered $\rm Mn^{3+}$ and $\rm Mn^{4+}$ ions may
not be accurate as the actual charge disproportionality of the Mn
atoms is much smaller than unity \cite{daoud-aladine02,thomas04,coey04}. 
The interplay of Mn charges, orbitals and spins would then be reduced
to the competition of the kinetic energy of the electrons and the
magnetic exchange interactions \cite{brink99,solovyev99}.  This
competition combined with the orbital degeneracy should lead 
to the experimentally observed CO-OO and drive the formation of 
1D ferromagnetic zigzag chains that are first loosely correlated and 
then lock in the magnetic CE structure of Fig.\ 1(a) below $T_N$. 
The idea that the CO-OO is independent of the magnetic interactions
seems to be far from true, and it is even possible that the CO-OO
is driven by the symmetry breaking caused by the magnetic
texture \cite{solovyev99,solovyev03}.  
It is remarkable that
despite the reported small charge disporportionality between the
Mn ions \cite{daoud-aladine02,thomas04,coey04} there are
dramatically distinctive behaviors of the Mn ions located at the
$\rm Mn^{3+}$ and $\rm Mn^{4+}$ sites as illustrated in Fig.\
\ref{fig:fig2}. The highly anisotropic and low-dimensional
behavior of the $\rm Mn^{3+}$ ions in this basically cubic
material  highlights the
role of the orbitals in the formation of the 1D spin chains in
this regime. 
Recently, Dagotto \cite{dagotto05} and Milward {\it et al.}\
\cite{milward05} have discussed the complex electronic behaviors
deriving from the strong energy competitions in the CMR
manganites. These authors proposed the presence of textured
electronically soft phases with incommensurate, inhomogeneous
and mixed orders in the paramagnetic state. The observed 
smectic liquid-crystal-like weakly coupled, 1D FM zigzag chains in
PCSMO confirm the presence of the sought after textured electronic
soft phase predicted in the manganites \cite{dagotto05,milward05} and 
cuprates \cite{kivelson98}.


In summary, we have studied the CO-OO and magnetic ordering in
the nearly half-doped manganite PCSMO using neutron scattering.
In the paramagnetic state $T_{N}<T<T_{CO}$, we find the
coexistence of FM, CO-OO and AF fluctuations. This coexistence arises
from the competition between the kinetic energy 
and the magnetic exchange, and from the orbital degeneracy of the Mn ions. 
While the CO-OO and AF fluctuations are strongly correlated in this temperature
region, they also exhibit distinctive behaviors.  Remarkably, the
AF short-range spin fluctuations of the $ \rm Mn^{3+}$ spins are
highly anisotropic and exhibit anisotropic smectic-like spin
texture, consistent with the presence of a electronically soft
phase as recently proposed in the literature. 


ORNL is managed by UT/Battelle, LLC for the US DOE under contract 
DE-AC05-00OR22725. This work was also supported by the US 
NSF DMR-0453804 and DOE and DOE-FG02-05ER46202. 
This work was performed under the US-Japan Cooperative Program 
on Neutron Scattering.


\begin{references}
\bibitem{smectic} We use the term smectic phase to describe the
observed anisotropic AF spin fluctuations although we do not know
the correlation of such fluctuations perpendicular to the
scattering plane. Strictly speaking, a liquid-crystal smectic
phase is not only orientationally aligned but also forms a
layered structure.
\bibitem{correlation} The correlation lengths are calculated using
$\xi=1/\kappa$, where $\kappa$ is the intrinsic half-width at half
maximum (HWHM) of the lineshapes after resolution corrections.
\end{references}
\end{document}